\documentclass[preprint,aps,prl]{revtex4}
 \usepackage{graphicx,epsfig}

\def\Tr{{\rm Tr}}
\def\--{\negthinspace - \negthinspace}

\def\ocite{\onlinecite}

\def\4ce{CeO$_2$}
\def\3ce{Ce$_2$O$_3$}
\def\xce{CeO$_{2-x}$}

\begin{document}

\title{Electron localization in pure and defective ceria by a unified LDA+U
approach}
\author{Stefano Fabris, Stefano de Gironcoli, and Stefano Baroni} 
\affiliation{SISSA and INFM DEMOCRITOS National Simulation Center,
  Via Beirut 2-4, I-34014 Trieste, Italy} 
\date{\today} 
\begin{center}
  \begin{abstract} 

 The electronic and structural properties of pure and defective cerium
 oxide are investigated using a unified LDA+U approach which allows to
 treat the different valence states of Ce occurring for different
 stoichiometries on a same ground, without any {\it a priori} assumptions
 on the defect chemistry.  The method correctly predicts the atomistic
 and electronic structures of the \4ce and \3ce bulk phases, as well as
 the subtle defect chemistry in reduced materials \xce, which controls
 the oxygen storage functionality of cerium-based oxides. The analysis of
 the energetics highlights the limits of the LDA+U method and possible
 extensions are proposed.  
\end{abstract}
\pacs{71.15.Mb, 61.72.Ji, 82.47.Ed}

\end{center}

\maketitle

Ceria based materials are among the key components of many advanced
catalysts used to produce hydrogen and to reduce air pollution.  The actual
catalyst is a complex device which couples the high chemical activity of
noble metals to the thermo-chemical stability of oxide substrates. Besides
providing a resistant support for the metal, ceria-based substrates take an
active role in the catalytic reaction by controlling the oxygen partial
pressure at the reaction sites, acting effectively as oxygen
reservoirs. The accepted qualitative mechanism of this phenomenon is based
on the following defect chemistry. A reduction of the oxygen partial
pressure promotes the release from the crystal lattice of atomic oxygen
which, after diffusing to the surface, recombines and desorbs as O$_2$
gas. The process leaves a charged vacancy in the lattice, V$_{\rm O}^{\cdot
\cdot}$, which is neutralized by the valence change Ce$^{4+}$ $\rightarrow$
Ce$^{3+}$ of two cations (substitutional Ce$_{\rm Ce}^\prime$ defects). The
release of one oxygen atom from the substrate drives therefore a
fundamental change in the electronic structure, leaving two electrons in
$f$-like empty states. Electronic correlation, due to the strong
localization of these states, have the effect of splitting the $f$ band
upon occupation, resulting in a fully occupied gap state at 1.5 eV above
the top of the valence band~\cite{Henderson03,Mullins99,Pfau94}. As a
consequence, reduced oxides, \xce, all result to be insulating.  The
electron population of this state is directly correlated to the number of
Ce$^{3+}$ ions, and it is employed as an estimator of the concentration of
oxygen vacancies~\cite{Henderson03}.

The localization of electrons on atomic-like $f$ states involves strong
correlation effects which are not captured by the most standard
implementations of density functional theory (DFT).  As a consequence of
this, existing first-principles
calculations~\cite{Skorodumova01,Skorodumova02,Gennard99,Conesa03} fail
to provide a unified picture of the different oxidation states of Ce
occurring in ceria. As a matter of facts, all of them adopt either one of
the following two assumptions, which are individually appropriate to
different oxidation states of Ce, but conflicting with one another: $i)$
in the core-state model (CSM) Ce $f$-states are treated as part of the
core, and therefore their contribution to the bonding process is totally
neglected~\cite{Skorodumova01}; $ii)$ the valence-band model (VBM),
instead, treats Ce $f$ electrons explicitly as valence electrons which
are therefore allowed to contribute to the chemical
bond~\cite{Skorodumova01,Gennard99,Conesa03}. The former approach
provides good structural properties for \3ce but not for \4ce (see
Table~\ref{tab-strct}), and it describes the oxide as an insulator {\it
by construction}. Most importantly its predictive power is limited: the
distribution of Ce$^{4+}$ or Ce$^{3+}$ ions has to be assumed as an input
of the calculation. As a consequence, the defect chemistry, which we have
seen to be determined by the valence change of Ce atoms, is difficult to
access.  On the opposite side, the VBM leads to good structural
properties for \4ce but not for \3ce (Table~\ref{tab-strct}), and
describes this latter structure as a metal. Neither of the two approaches
predicts the existence of the gap state experimentally observed to occur
in partially reduced ceria \xce, therefore failing to describe the
localization-delocalization transition of $f$ states, which is at the
basis of the oxygen-storage mechanism.

In the present work these difficulties are overcome using an LDA+U approach
which allows to describe all the Ce atoms within a single, unified,
approximation regardless of their oxidation state.  This is particularly
important in the study of complex, heterogeneous, systems---such as {\it
e.g.} a molecule adsorbed on an oxide-supported metal nano-particle---where
the oxidation state of each Ce atom is not known in advance. Electron
correlations within Ce $f$ states are modeled by a Hubbard $U$ contribution
added to the traditional density-functional self-consistent potential. The
method describes explicitly the $f$ states and is highly predictive: it
correctly describes the electronic structure of pure and defected cerias
(always predicted to be insulators); it captures the valence change due to
non-stoichiometry (and the related features in the electronic structure);
and it allows the intrinsic defects Ce$_{\rm Ce}^\prime$ to occupy the most
energetically favorable position with respect to V$_{\rm O}^{\cdot \cdot}$
without any {\it a priori} assumptions on their relative geometry.
Moreover, it allows to estimate the formation energetics for pure and
defective ceria phases, which turn out to be consistent with experiments.

The results presented in this work were obtained by ab initio DFT-based
calculations employing the local density approximation for the exchange
and correlation potential, and were performed with the PWscf
package~\cite{BaroniPW}. The crystal valence wave functions were
represented by a plane-waves basis set limited by the kinetic energy of
30 Ry. Their interactions with nuclei and core electrons were described
by non-local ultrasoft psudopotentials~\cite{Vanderbilt90} constructed
for the following atomic configurations: O $2s^2 2p^4$, and Ce $5s^2 5p^6
6s^2 5d^1 4f^1$. The Hubbard $U$ contribution to the energy functional
was treated after the formulation of Cococcioni and de
Gironcoli~\cite{CococcioniTh,Cococcioni03}. The on-site parameter $U$ for
Ce was fixed to 3 eV, following the analysis of Ref.~\ocite{BalducciUnp}
where it was estimated with a variational method to lye between 3 and 3.5
eV. Two concentrations of vacancies were considered in the analysis of
defective cerias \xce: $x=0.125$ and $x=0.03125$. The defects were
modeled with the supercell method by removing one oxygen atom from the
supercells containing, respectively, 24 and 96 atomic sites.

The LDA+U energy functional reads:
\begin{equation}
    E_{LDA+U} = E_{LDA} + \frac{U}{2} \sum_I \Tr\left ( {\bf n}^I
      ({\bf 1}-{\bf n}^I ) \right ), \label{eq:LDA+U}
\end{equation}
where the ${\bf n}^I $'s are $M\times M$ matrices ($M$ being the
degeneracy of the localized atomic orbital, $M=14$ in the case of $f$
orbitals), projections of the one-electron density matrix,
$\widehat\rho$, over the $f$ manifold localized at lattice site $I$:
\begin{equation} 
  \langle \phi^I_{m\sigma}| \widehat\rho | \phi^I_{m'\sigma'} \rangle = 
  n^{I\sigma}_{mm'} \delta_{\sigma\sigma'}.    \label{ns}
\end{equation} 
In the above equation $\phi_{m\sigma}$ denotes a localized (spin)
orbital, with angular and spin quantum numbers $m$ and $\sigma$, and the
orthonormality over the spin variables is a consequence of the
collinearity of the spin structure. The second term in
Eq. (\ref{eq:LDA+U}), which we call $E_U$, is positive definite for $U>0$
because the eigenvalues of the ${\bf n}^I$ matrices---{\it i.e.} the
occupation numbers of the $f$ orbitals---lay in the range $[0,1]$. Note
that, as soon as $U$ is large enough to open a gap between occupied and
unoccupied $f$ states, the effect of $E_U$ is to favor the $f$ occupation
numbers to be close to either 0 or 1. At these extremes, $E_U$
is strictly 0 and the total energy is back to the LDA value.

Ceria is known to exist in the cubic fluorite-type \4ce ($Fm3m$) and the
hexagonal $A$-type \3ce ($P\bar{3}m1$) bulk phases. The latter can be
considered as a highly defective \4ce structure where one every four oxygen
atoms is missing, with the vacancies being ordered along non intersecting
$\langle 111 \rangle$ directions, and with all the Ce ions nominally in the
3+ valence state~\cite{Skorodumova02}. The \3ce structure is
anti-ferromagnetic~\cite{Pinto82}, therefore the calculations for this
phase were carried out in the local-spin-density approximation.

\begin{table}[floatfix]
\caption{Calculated structural properties of the bulk \4ce and \3ce phases
compared with previous theoretical values in the valence-band and
core-state models (VBM and CSM), and with experimental measurements.}
\label{tab-strct}
\begin{center}
\begin{tabular}{l|c|c|c|c|c}
     &  \multicolumn{2}{c}{\4ce} & \multicolumn{3}{|c}{\3ce} \\ 
     & $a_0$ [\AA\ ] & $B$ [GPa] & $a_0$ [\AA\ ] & $c/a$ & $B$ [GPa] \\ \hline
LDA VBM Ref.~\onlinecite{Skorodumova02} & 5.39 & 214.7 & 3.72 & 1.55 & 208.6\\
LDA CSM Ref.~\onlinecite{Skorodumova02} & 5.56 & 144.9 & 3.89 & 1.55 & 165.8\\
LDA+U This work                       & 5.41 & 210.6 & 3.81 & 1.55 & 149.8\\
Expt.                               & 5.41 & 204-236 & 3.88 & 1.55 & -- \\
\end{tabular}
\end{center}
\end{table}

The calculated structural properties for the bulk polymorphs are
summarized in Table~\ref{tab-strct} and are in good agreement with
experiments. The Hubbard contribution to the energy functional increases
the lattice parameter by 0.4\% in \4ce and by 2.5\% in \3ce with respect
to the LDA values. Note how the LDA+U approach provides a good prediction
of the structural properties for both polymorphs, therefore combining the
achievements of the VBM for \4ce with the ones of the CSM for \3ce.

\begin{figure}[floatfix]
\caption{Density of electronic states for pure \4ce (a) and \3ce (b) bulk
 phases, and for defective (c) ceria structures \xce. Occupied states
 are indicated as shaded areas, and the zero
 energy is set to the top of the valence band.}
\label{fig-dos}
 \centerline{\psfig{file=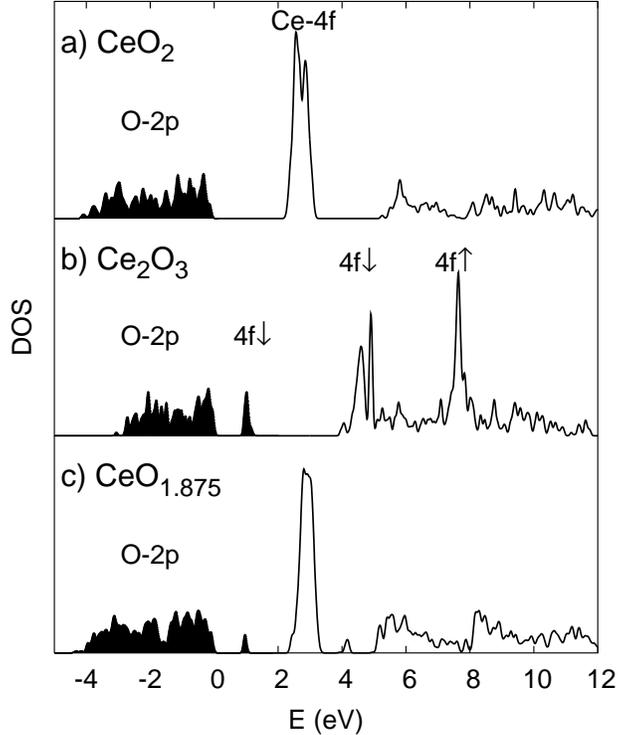,width=8cm,angle=0}} 
\end{figure}

The density of electronic states (DOS) for the two polymorphs calculated
at the equilibrium lattice parameters are shown in
Figure~\ref{fig-dos}. Occupied states are indicated by shaded areas, and
the zero energy is the top of the upper valence band (with prevalent
O-$2p$ character). For \4ce (Figure~\ref{fig-dos}a), the sharp band
centered at $\approx$ 3 eV is formed by fairly localized atomic-like
Ce-$4f$ empty orbitals.  The onset of the conduction band is 5.1 eV above
the highest occupied state, exhibiting the typical LDA underestimate with
respect to the measured gap of 6.0 eV~\cite{Wuilloud84} (following
Ref.~[14] we do not consider the $f$-manifold as conduction states). The
calculated electronic structure for \3ce is in good agreement with
experiments: it predicts an anti-ferromagnetic insulator, with a magnetic
moment of 2.25 $\mu_{\rm B}$/molec (the experimental value is 2.17
$\mu_{\rm B}$/molec~\cite{Pinto82}). The corresponding DOS is displayed
in Figure~\ref{fig-dos}b, showing a splitting between occupied and
unoccupied states within the $f$ manifold.  Due to the effects of the
Hubbard energy term, the Ce-$4f$ states do not form one single band ---
as in the \4ce structure or in previous VBM calculations --- but are
split in three energy regions: a doubly occupied state at $\approx 1.2$
eV above the top of the valence O-$2p$ band and corresponding to
electrons localized on the Ce$^{3+}$ atoms (with alternating spins along
the $z$ axis). The position and intensity of this peak (relative to the
O-$2p$ band) are in excellent agreement with experimental photoemission
measurements~\cite{Mullins99}. In the range of U suggested by
Ref.~\cite{BalducciUnp} (3-3.5 eV), the energy of the gap state depends
linearly on U spanning the values between 1.2 and 0.9 eV (with respect to
the top of the valence band). The remaining 13 $f$ states are also split
in two bands centered at 4.5 and 7.5 eV, and grouping the 6 and 7 states
having their spin parallel and anti-parallel to the one of the electron
localized on their respective site.

\begin{figure}[floatfix]
\caption{Isosurface in the charge density (grey)
  corresponding to the occupied gap-state in CeO$_{1.969}$ (left) and \3ce
  (right). Black and light grey circles  denote Ce and O atoms.}
\label{fig-ce2o3}
\begin{tabular}{lcr}
\parbox{5.5cm}{\large a) CeO$_{1.969}$}  &
\parbox{3.1cm}{\large b) \3ce} \\
\parbox{5.5cm}{ \psfig{file=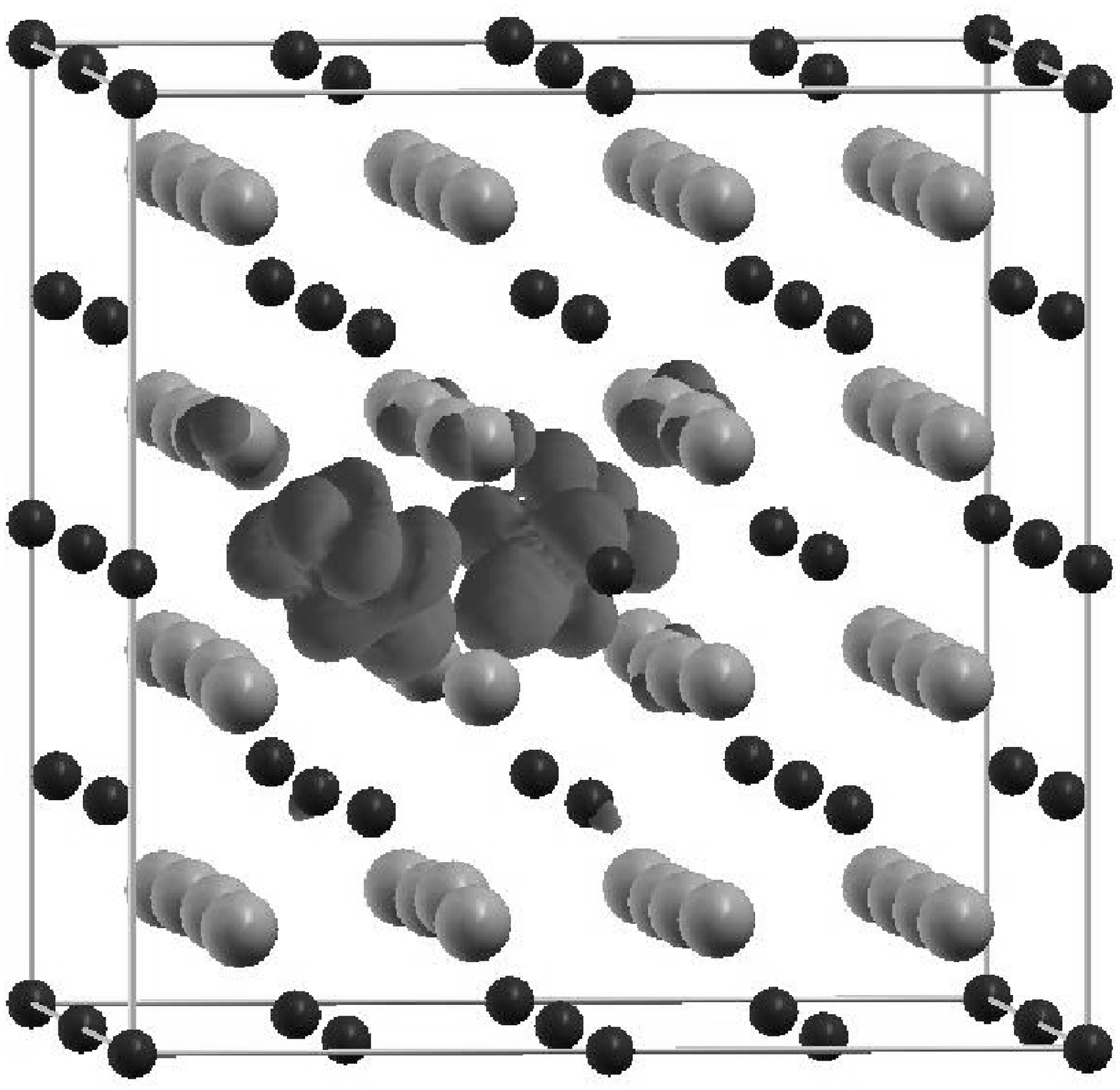,width=5.3cm,angle=0} \hfill } &
\parbox{3.1cm}{ \hfill \psfig{file=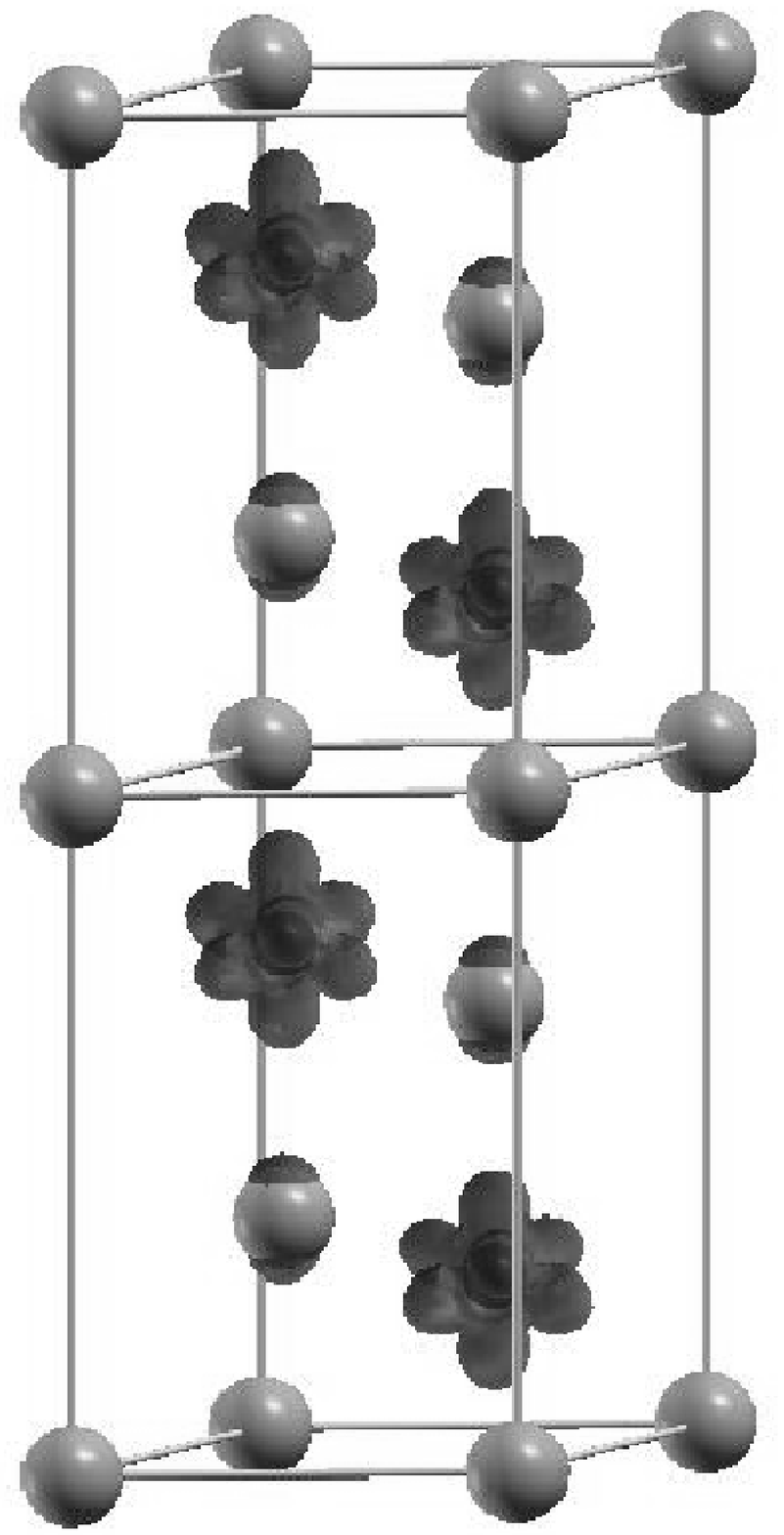,width=2.8cm,angle=0} } 
\end{tabular}
\end{figure}

An oxygen vacancy is formed by removing a neutral oxygen atom from a \4ce
supercell. As a consequence, two electrons have to be accommodated in the
$f$ states above the Fermi energy of Figure~\ref{fig-dos}a. Our LDA+U
model predicts the localization of two electrons on two Ce atoms
neighboring the vacancy. This has the effect of formally modifying the
valency of Ce from $+4$ to $+3$. This result is in agreement with the
experimental evidence that the compensating defects are Ce$_{\rm
Ce}^\prime$ which tend to cluster around the oxygen vacancy, forming a
defect aggregate. An electronic configuration in which two electrons were
initially localized on Ce atoms far from the vacancy led to the same
self-consistent solution with the Ce nearest neighbors to the vacancy
being formally Ce$^{3+}$. Upon removal of an oxygen atom, the O
atoms neighbouring the vacancy relax toward the vacancy by $\approx$
0.125 \AA, and the Ce atoms outward by $\approx$ 0.08 \AA. Moreover, the
O atoms neighbouring the Ce$_{\rm Ce}^\prime$ relax away from these
defects (see our previous description on the supercells used to model these
systems). The DOS for a model of defective ceria, CeO$_{2-x}$, is shown
in Figure~\ref{fig-dos}c for $x=0.125$. One sees that the DOS displays in
this case features which are intermediate between those of the \3ce and
\4ce structures: the gap state at $\approx$ 1.2 eV (due to occupied $f$
states on Ce$^{3+}$ atoms as in \3ce), and the sharp unoccupied $f$ band
of the Ce$^{4+}$ atoms at 3 eV. The position and relative weight of the
peak corresponding to the occupied gap state are in excellent agreement
with observed photoemission spectra~\cite{Henderson03,Mullins99,Pfau94}.
In Figure~\ref{fig-ce2o3}a we display the integrated charge density
corresponding to the gap state in CeO$_{2-x}$. This figure demonstrates
that, in these dilute regimes, the effects of the formation of a vacancy
on the electronic structure are limited to the first shell of atoms
around the defects (vacancy and compensating Ce$^{3+}$).

Let us now examine the energetics of reduction of ceria, and focus in
particular on the two reactions: the \4ce--\3ce transition, 2\4ce
$\rightarrow$ \3ce + 1/2O$_2$(g) (whose enthalpy measured at 298 K is
$-0.58$ Ry~\cite{CRCceo2}), and the formation of a vacancy (whose
experimental heat of reduction is $0.34\-- 0.37$
Ry~\cite{Tuller75,Chang88}). The LDA+U approach severely underestimate
the energetics of these reactions: $-0.33$ Ry for the \3ce-\4ce
transition, and 0.26 Ry for the vacancy formation. This error is not due
to the local density approximation itself, since gradient corrected
calculations (including the Hubbard term) also led to quite similar
underestimates.

The source of this error can be traced back to some details of the
implementation of the LDA+U method which are best evidenced by
analyzing the occupation matrices, Eq.~(\ref{ns}). The integrated charge
of the gap state is exactly 2 (one electron for each Ce$^{3+}$
atom).  Inspection of the isosurface charge density displayed in
Figures~\ref{fig-ce2o3} shows that the the gap state is not localized
exclusively on Ce$^{+3}$ atoms but has small projections also on the
first neighbouring O sites. As a consequence, when this state is
projected onto Ce-$f$ orbitals, Eq.~(\ref{ns}), the resulting occupation
number is clearly smaller than 1 (in fact, $\approx 0.98$).  The problem
is even more clear in \4ce, where, due to the prevalent ionic bonding,
the Ce-$4f$ states should be empty and the occupancies very close to
0. This is not the case, since most of the calculated occupancies are
small, but not vanishing (in fact,$\approx 0.01\-- 0.02$). In the
particular case of the $xyz$ component, which points along the Ce-O bond,
the occupancy is much larger ($\approx 0.2$). This large value comes from
the partial covalent character of the Ce-O bond, has nothing to do with
correlation, and should not contribute to the Hubbard energy, $E_U$.

We conclude that the errors in the vacancy formation energy has to be
traced back to the spurious occupancies resulting from the particular
choice of the projector in~(\ref{ns}), which picks up bonding charge from
neighbouring O atoms. Nevertheless, in this relatively simple case (where
the occupied $f$ state is well separated from the valence band) a simple
and effective remedy can be found.  If the states used to calculate the
occupation matrices, Eq. (\ref{ns}), instead of being the atomic Ce-$f$
states, were those resulting {\em self-consistently} from the calculation
({\em i.e.} the Wannier functions of prevalent Ce-$f$ character), then
the occupancy corresponding to the gap state would be one, and that
corresponding to other orbitals of prevalent Ce-$f$ character zero, {\em
by construction}. This argument shows that---in the present situation
where the split state is well separated from all the others---a better
characterization of the localized orbitals which define $E_U$ would
simply result in a vanishing contribution of $E_U$ to the final,
self-consistent, energy. A quick-and-dirt fixup to the vacancy formation
energy problem seems therefore to be to use $E_U$ to generate the correct
electronic structure, but to ignore it altogether in the calculation of
the total energy.  This amounts to use the LDA+U model to generate the
molecular orbitals, and revert to the LDA functional for the calculation
of energies.  With this extremely simple scheme, also the energetics of
the two reactions result in good agreement with experiments: $-0.610$ Ry
for the \3ce-\4ce transition (experimental value is
-0.58Ry~\cite{CRCceo2}), and 0.361 Ry for the vacancy formation
(experimental values are $0.34\-- 0.37$ Ry~\cite{Tuller75,Chang88}).

We stress that the dependence of the results on the projector used to
define the occupancies is important on the energetics, but have little
effect on the electronic structure. The atomistic and electronic
structures of pure and defective cerias, as well as their energetics, can
therefore be captured by a {\it modified} LDA+U method which takes
advantage of the Hubbard $U$ effect (essential for giving the correct
description of the electrons) in the self-consistent electronic problem,
and whose energy has been ``corrected'' a posteriori to account for the
spurious occupancies due to partially covalent character of the chemical
bond.

  In summary, the LDA+U method provides a simple and accurate way to
  account for the subtle mechanism of $f$-electron
  localization-delocalization which determines the mixed valence
  character of Ce and which is at the basis of the oxygen storage
  properties of ceria. This method is accurate and predictive, opening
  the way to the {\em ab-initio} study of complex structures of
  technological interest---such as {\em e.g.} ceria-supported metal
  nanoparticles---which cannot be easily treated by more empirical
  approaches where the valence state of individual Ce atoms is
  pre-assigned. Limitations of this approach show in the calculation
  of reduction/oxidation energies, which can however be fixed in the
  present case by a simple prescription on the energy functional.

\begin{acknowledgments}
We wish to thank J. Kaspar for prompting our interest in this subject and
for fruitful discussions. We are also grateful to G. Balducci for many
fruitful discussions and for allowing us to use data from
Ref.~\ocite{BalducciUnp} prior to publication. Calculations have been
made possible by the SISSA-CINECA scientific agreement and by the
allocation of computer resources from INFM Progetto Calcolo Parallelo
Computer. Molecular graphics has been generated with the XCrySDen
software~\cite{KokalijXC}.
\end{acknowledgments}

\end{document}